\documentclass[aps,prl,twocolumn,groupedaddress,showpacs]{revtex4}
\usepackage{graphicx}
\usepackage{latexsym}
\begin{document}

\title{Anomalous Coulomb drag in electron-hole bilayers}

\author{A.F. Croxall}\author{K. Das Gupta}\email{kd241@cam.ac.uk}\author{C.A. Nicoll} \author{M. Thangaraj}
\author{H.E. Beere} \author{I. Farrer} \author{D.A. Ritchie} \author{M. Pepper}
\affiliation{Cavendish Laboratory, University of Cambridge, J.J. Thomson Avenue, Cambridge CB3 0HE, UK.}

\begin{abstract}
We report Coulomb drag measurements on GaAs-AlGaAs electron-hole bilayers.
The two layers are separated by a 10 or 25nm barrier. Below T$\approx$1K we find two features that a Fermi-liquid picture cannot explain. First, the drag on the hole layer shows an upturn, which may be followed by a downturn. Second, the effect is either absent or much weaker in the electron layer, even though the measurements are within the linear response regime. Correlated phases have been anticipated in these, but  surprisingly, the experimental results appear to contradict Onsager's reciprocity theorem.
\end{abstract}

 \maketitle
%\date{\today}

\begin{figure}[h]
\includegraphics[width=8.75cm,clip]{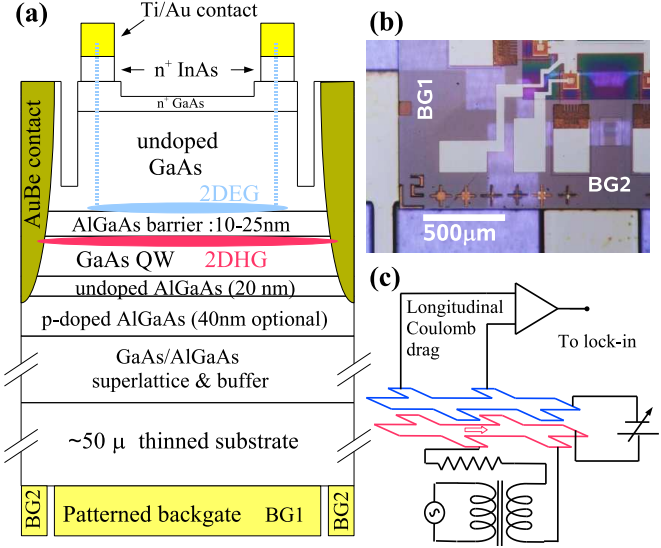}
\caption{\label{sample-schematic} (Color online) (a) Generalised
schematic of the devices. The p-doping and  hole QW width vary between
devices, as given in table \ref{table1}. (b) Photograph of the sample showing the alignment of the backgates.
(c) Schematic of the  circuit for Coulomb drag measurements. The backgate and the topgate
are omitted for clarity.}
\end{figure}

%\section{Introduction}
Pairing between quasiparticles constituting a Fermi liquid system
leads to some of the most interesting phenomena in solid-state
physics, such as paired atoms in superfluid $^3$He and Cooper pairs of
electrons in superconductors. The presence of electrons and holes in a
semiconductor naturally leads
to the possibility of binding like that found in a hydrogen
atom. The bosonic nature of these excitons follows, because they are paired states of spin-$\frac{1}{2}$particles.
The very short lifetimes (usually nanoseconds) and the charge neutrality
of these bound pairs place them outside the realm of transport measurements.
Short lifetimes may also inhibit the formation of coherent equilibrium
phases like a Bose condensate\cite{blatt}. Separating the
electrons and holes spatially, with a thin barrier, would prevent
recombination and lead to increased lifetimes. Exciting predictions have been made on the possibility of novel phases in such electron-hole (EH) bilayers. Early proposals\cite{lozovik} relied on n-semiconductor-insulator-p-semiconductor structures to
achieve this. The rapid improvements in  GaAs/AlGaAs heterostructure technology in the 1980's and subsequent development of closely spaced double
quantum-well  structures in the 1990's led to the first realistic
possibilities of making such a system. The Coulomb drag technique,
in which a current passed through one layer induces an open circuit
voltage in the other layer, is a direct measure of the interlayer
interaction.  The effect\cite{price,pogrebinsky}, analogous to momentum
transfer between layers of a viscous fluid, was first experimentally demonstrated between a pair of 2-dimensional electron gases (2$\times$2DEG)\cite{gramilla}.  In EH bilayers a  divergence of longitudinal
Coulomb drag  was predicted to occur at the onset of an excitonic condensation\cite{vignale,hu}. An excitonic dipolar superfluid with a phase that couples to the gradient of the vector potential has also been conjectured\cite{balatsky,joglekar}. However, a recent careful calculation\cite{dassarma} of the electron and hole polarizabilities  (in an EH bilayer) has  emphasized the fact that a finite interlayer scattering rate ({\it i.e.} drag) at $T=0$ is not possible within the Fermi liquid picture. The $\nu$=1 bilayer state in 2$\times$2DEG and 2$\times$2DHG emulates a true EH bilayer in certain ways\cite{eisenstein2}. Experiments on these systems showed a remarkable collapse of the Hall voltage and finite drag as T$\rightarrow$0, suggesting transport by charge neutral entities\cite{kellog}.\\
The fabrication of closely spaced and independently
contacted EH bilayers presents considerable difficulties
compared to electron-electron and hole-hole bilayers. These are now
well understood \cite{keogh} and significant improvements have been
made \cite{keogh, pohlt, seamons, backgate} since the first reported device
by Sivan {\it et al} \cite{sivan}. Fabrication of EH bilayer devices where the barrier between electron and hole layers
is similar to the excitonic Bohr radius of GaAs ($\approx$
12nm) and measurement of Coulomb drag down to milliKelvin
temperatures are now possible.\\
%\section{Devices}
In this Letter we report Coulomb drag data from four  devices (table \ref{table1}).
Generalised structure of these is shown in Fig.\ref{sample-schematic}.
The details of the wafer design, growth, bandstructure and processing techniques have been described earlier \cite{keogh, backgate}. We
start with an inverted 2-dimensional hole gas (2DHG) with little or no doping, such that it can be  backgated after the sample is thinned to about
50 $\mu$m. Using the 2DHG as a gate we induce a 2DEG
above an  AlGaAs barrier. The 2DEG forms only under an interlayer
bias, slightly higher than the bandgap of GaAs,
1.52V. The contacts to the 2DEG must not penetrate the barrier,
to avoid leakage current between the two gases.
This is achieved by using the "negative Schottky barrier" at an n+
InAs/metal interface\cite{keogh}, which requires no annealing.
A near-flatband condition must be maintained between the InAs/GaAs and the
2DEG for the mechanism to work. The electron density ($n$) is fixed by the
interlayer bias ($V_{eh}$) only. The hole density ($p$) is a function of
$V_{eh}$ and the backgate voltage ($V_{bg}$). By measuring $n$ and $p$ at different $V_{eh}$ but fixed $V_{bg}$, we can obtain  a quantitative measure of the interlayer capacitance and hence the peak-to-peak separation ($d$) of the wavefunctions. For the 25nm barrier, this gives $d\approx$37nm and for the 10nm barrier  $d\approx$25nm.

\begin{figure}[h]
\includegraphics[width=7.8cm,clip]{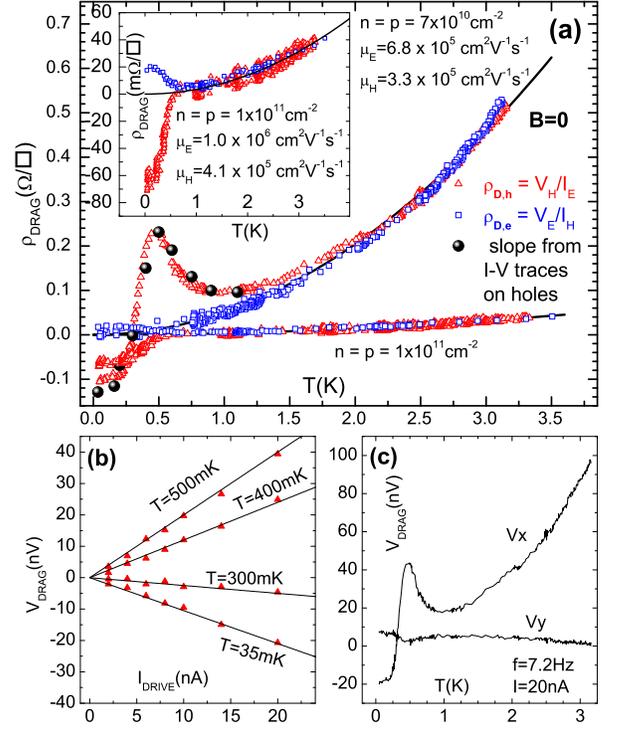}
\caption{\label{b135c3-4drag} (Color online) (a)Data from device A. The drive current was 20nA, 7Hz for all the traces.
The solid black line are best-fits to a $T^2$ behaviour. The inset shows an expanded view of the lower trace at n=p=$1.0\times10^{11}$cm$^{-2}$. The upturn was no longer observed at this density on the holes, but the downturn can be
seen. 
(b) Raw data showing the linearity of  drag voltage with drive current and (c) relatively flat behaviour of the out of phase component($V_y$) as measured by the lock-in, when the in-phase ($V_x$) component goes through an upturn
and sign reversal.}
\end{figure}

\begin{figure}[t]
\includegraphics[width=8.7cm,clip]{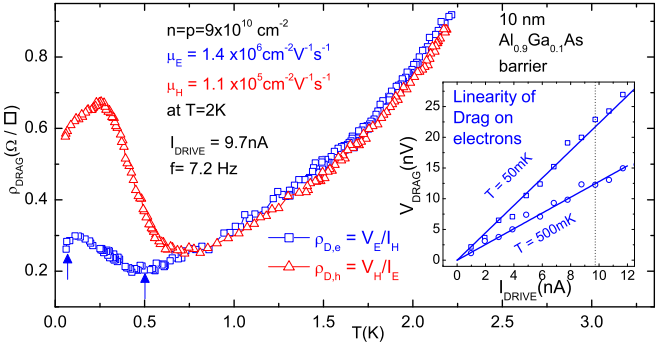}
\caption{\label{b138c4-1mx400}(Color online) Data from device D.
The upturn and the beginning of  the downturn are clearly seen. A much weaker effect is seen for the electrons than the holes. The inset shows that the measurements are clearly within the linear response regime.}
\end{figure}

Devices A and D, where matched densities were obtained were measured
till $\sim$50mK. The temperature dependence of the Coulomb drag on the hole layer showed an upturn as the temperature was lowered below T$\approx$1K, followed by a downturn (see Figs. \ref{b135c3-4drag}a \& \ref{b138c4-1mx400}). Devices B \& C were measured in the range 6K-300mK, and showed an upturn. We emphasize that these features are only
seen in devices with very low barrier leakage - typically $I_{leak}
<$ 50pA over a Hall bar 600$\mu{m}$ $\times$ 60$\mu{m}$ in
size at an interlayer bias of $\sim$1.6V.
A low temperature  upturn in Coulomb drag in similar density and temperature ranges has also recently been reported \cite{condmat08081322}.

\begin{figure}
\includegraphics[width=8.5cm,clip]{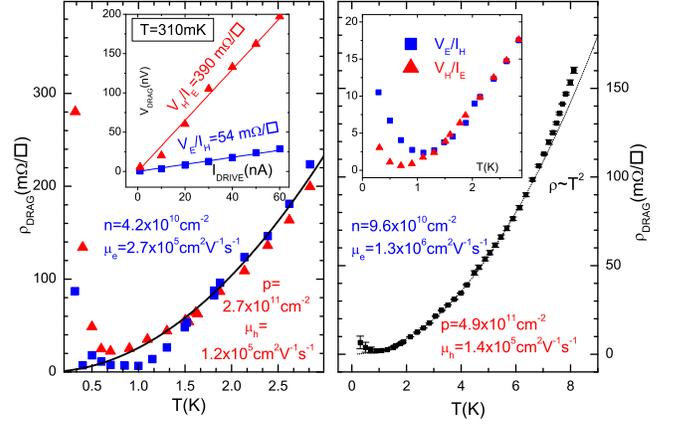}
\caption{\label{a4005andb89c2-4} (Color online) (Left) Data from
device B. The  inset shows that  the linearity of drag voltage and
drive current is not compromised.
(Right)Data from device C. In the main figure, the
error bars represent the difference of the drag voltage measured by
interchanging the drag and drive layers. The inset shows the
expanded view of the upturn.}
\end{figure}

\begin{table}[h]
\caption{\label{table1}Summary of the device parameters. Devices A,B
\& C have a 25nm Al$_{0.3}$Ga$_{0.7}$As  barrier and D has a 10nm
Al$_{0.9}$Ga$_{0.1}$As  barrier. %The doped devices have a 20nm undoped %spacer.
}
\begin{tabular}{|l|c|c|c|c|c|}
\hline \parbox{1.5cm}{device \\ {[}wafer ID{]}} &\parbox{2.5cm}{
p-doping\\(cm$^{-3}$)}&\parbox{1cm}{\vspace*{10pt}QW\\width\vspace*{10pt}} &
barrier &\parbox{1.5cm}{ matched densities possible} \\
\hline
A [A4142] & undoped & 20nm & 25nm & Yes\\
\hline %&&&\\
B [A4005] & \parbox{2.75cm}{$1\times10^{17}$(Carbon)}  & 40nm & 25nm & No\\
\hline %&&&\\
C [A3524] & \parbox{2.75cm}{$2\times10^{18}$(Beryllium)} & 40nm & 25nm & No\\
\hline
D [A4268] & \parbox{2.75cm}{$5\times10^{16}$(Carbon)} & 20nm & 10nm & Yes\\
\hline
\end{tabular}
\end{table}
%\section{Results}
Fig.\ref{b135c3-4drag} shows the measurement of Coulomb drag using device
A in full detail.  The drag voltage was measured in two ways, by sending current through the electrons  and measuring the open-circuit voltage
across the holes ($\rho_{D,h}=V_h/I_e$) or  by sending
current through the holes and measuring the voltage across the
electrons ($\rho_{D,e}=V_e/I_h$). As long as the current is low enough so that the system is in the linear response regime,  thermodynamic
arguments\cite{casimir} predict that $\rho_{D,e}$ = $\rho_{D,h}$.
In our data this is well satisfied above T$\sim$1K and
$\rho_{D,e/h} \propto T^2$ approximately. The origin of this  behaviour  is well known\cite{jauho}. Below T$\sim$1K, as T decreases, $\rho_{D,h}$  starts increasing, passes through a maximum, and decreases in both devices A (Fig.\ref{b135c3-4drag})\& D(Fig.\ref{b138c4-1mx400}). It does not appear to be going to zero in either case, as $T\rightarrow$0. A finite drag resistivity, at $T=0$ is not possible within a  Fermi liquid picture. It has only been predicted for a paired electron-hole superfluid \cite{vignale,hu} and an incompressible paired Quantum Hall state, with the temperature dependence (near T=0) determined by disorder and not only the available phase space\cite{zhou-kim}. In this context, it is important to point out that if the scattering rate is calculated  using the Born approximation, the square of the interlayer screened Coloumb potential is used and the matrix element will not distinguish an attractive interaction from a repulsive one. Emergence of binding or pairing, that may lead to a qualitative change in the matrix element for the interlayer scattering rate, cannot not be correctly accounted for by simple first order theory of scattering \cite{hu}.
The initial upturn, that we observe, can be qualitatively explained if a small fraction of the particles enter into a paired state and has been anticipated \cite{vignale,hu}. However there are two rather surprising aspects of our data, which a simple pair formation hypothesis cannot explain. First,
the upturn appears to be followed by a downturn at low densities and low temperatures. Second, the effect is absent or much weaker, when the current is driven through the hole layer.
 Figs. \ref{b135c3-4drag}b and \ref{b138c4-1mx400}(inset) show that the system continues to be in the linear response regime throughout the temperature range where the upturn (in all the devices), downturn (in device A \& D) and the sign reversal (in device A only) are observed. The effect is free
from  thermal hysteresis or drifts with time - the data of
Fig.\ref{b135c3-4drag} represented by the red triangles and blue squares was
collected while the temperature was decreasing, the data points
represented by large black dots were collected while the device was
being warmed up.
Data from devices B \& C (measured
down to 300mK) show a similar upturn (see Fig.
\ref{a4005andb89c2-4}). The effect is very small in device C,
possibly because of the high hole density.
The presence of this upturn over a wide range of densities from matched to strongly unmatched, suggests that equal densities are not a necessary condition to observe this feature. We have verified by shifting the biasing point from one end of the hall bar to another, that leakage errors ($V_{error}\approx I_{leak}R_{single layer}$) do not change the measured values of drag significantly. Also devices with high leakage (due to defects in the barrier, gate leakage etc.) do not show any of  these low temperature features. Measurements using currents as low as 1nA did not reveal any deviation from linear response. Straight line fits to the $I_{drive}$-$V_{drag}$ data (Figs. \ref{b135c3-4drag}a, \ref{b138c4-1mx400}(inset) \& \ref{a4005andb89c2-4}(left, inset)) show zero offset indicating the absence of rectified noise in the system. If indeed a critical current exists, (as may happen for a  pinned lattice or a Wigner crystal state) below which the layers behave symmetrically, it appears to be extremely small. Since there is no change of slope in the $I_{drive}$-$V_{drag}$ traces at low currents, we rule out Joule heating as well.

\begin{figure}
\includegraphics[width=8.5cm,clip]{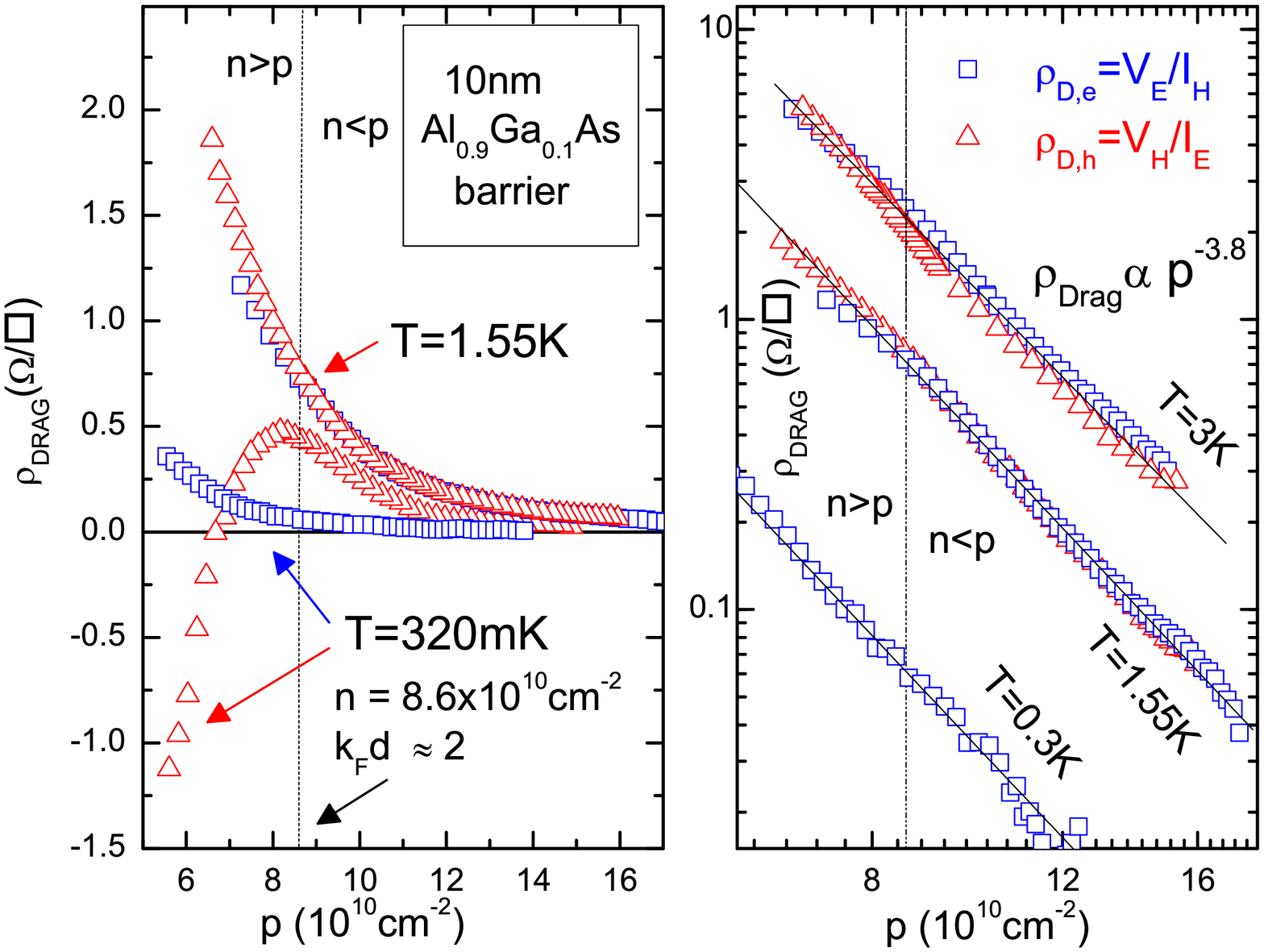}
\caption{\label{b138c4-1data}(Color online) Data from device D, with
a 10nm Al$_{0.9}$Ga$_{0.1}$As barrier.  The dotted vertical line in
both panels denote the point where $n=p$. At this density,
considering an effective peak-to-peak separation of $d=25$nm, we get
$d/l \approx 2$, with $l=1/\sqrt(2\pi n)$. The red triangles denote
drag measured on the holes ($\rho_{D,h}$), the blue squares denote
drag measured on electrons($\rho_{D,e}$). The data at T=1.55K is
repeated on both panels for comparison. }
\end{figure}
Fig.\ref{b138c4-1data} shows the effect of keeping the carrier density of
one layer constant, while varying the other for device D, which had a 10nm Al$_{0.9}$Ga$_{0.1}$As barrier. The right panel shows that a power law
$\rho_{D,e} \propto p^{-3.8}$ holds between 0.3-3K. The traces taken at T=3K,1.5K show that $\rho_{D,h}$ and $\rho_{D,e}$ agree very well, but not at 0.3K. Similarly, if  the hole density is constant, we get $\rho_{D,h} \propto n^{-0.5}$ (data not shown).  If we consider only $\rho_{D,e}$ then even at T=0.32K, the $n=p$ condition does not appear to be special. On the other hand, as observed in other devices, $\rho_{D,h}$ shows a strong deviation
from a simple power law ($\sim T^2$) at T$\sim$1K. At matched densities in bilayers, $d/l$ is the same as $k_Fd$. It is interesting to ask whether the nature of $\rho_{D,h}$ changes around $n=p$ or when large angle scattering starts dominating({\it i.e.} $k_Fd$ is small). However at this stage we do not have enough data to address this point.\\
Coulomb drag measurements on 2x2DEGs in a magnetic field have been reported by  several authors\cite{feng,lilly,lok,hill,muraki}. At large filling factors
({\textit i.e.} low magnetic fields), it is found that if the
deviation from half filling (of the highest Landau level) is
opposite in the drive and drag layers, then $\rho_{D}$ is
negative (electron-hole like) at low temperatures. However at higher
temperatures, when $k_{B}T$ becomes larger than disorder broadening of the levels, $\rho_{D}$ turns positive again.  The temperature above which
negative drag is no longer observed  is consistent with experimental
values of disorder broadening of the Landau levels in a sample,
($\sim$1K in high mobility samples,  estimated from
the quantum scattering times, $T\sim \hbar/k_{B}\tau$). It has also been
emphasized that the excitations near the Fermi surface must not have
particle-hole symmetry for $\rho_{D}$ to be non-zero\cite{muraki,feng}. If the
Fermi level in any one layer lies exactly at the centre of a
spin-resolved Landau level, then that level would acquire this
symmetry and its contribution to drag would diminish. As the Fermi
energy passes through successive levels, a complex sequence of
positive, zero and negative drag can result.
The data shown in Fig. \ref{b135c3-4drag}, has a strong resemblance to that reported in
\cite{muraki} for $\nu=7.5,9.5$ etc. This similarity is
unexpected at $B=0$, for the results presented here. Without a
mechanism of generating discrete levels (like Landau levels) the
concept of disorder broadening cannot be applied. In 2DEGs and
2DHGs, one expects continuous $E(k)$ dispersion.  Unless a gap
appears, level-broadening is not meaningful. One might
speculate about the appearance of a gap in a coupled
2DEG-2DHG system due to pairing or localization, but we refrain from
doing so at this point.\\
In conclusion, we  have shown that the interlayer scattering rate in closely spaced electron-hole bilayers exhibits novel features over a wide density range, which cannot be explained within the Fermi-liquid picture.  While the origin of these  is not understood at present, our experimental data is distinctly in the linear response regime and as such the disagreement with the reciprocity theorem \cite{casimir}, may point to a robust aspect of the bilayer ground state. \\
\textit{Acknowledgements.} We thank K. Muraki for useful suggestions and and J. Waldie for careful reading of the manuscript. The work was funded by EPSRC, UK. MT acknowledges support from the Gates Cambridge Trust.

\end{document}